# GEE-TGDR: A longitudinal feature selection algorithm and its application to lncRNA expression profiles for psoriasis patients treated with immune therapies

**Authors:** Suyan Tian[1§], Chi Wang[2,3] and Mayte Suarez-Farinas[4,5§]

[1]Division of Clinical Division, The First Hospital of Jilin University, Changchun, Jilin, China. 130021

[2]Department of Internal Medicine, College of Medicine, University of Kentucky, 800 Rose St., Lexington, KY 40536, USA

[3]Markey Cancer Center, University of Kentucky, 800 Rose St., Lexington, KY, 40536, USA

[4]Department of Population Health Science & Policy, The Icahn School of Medicine at Mount Sinai, New York City, NY, 10029, USA

[5]Department of Genetics and Genomics, The Icahn School of Medicine at Mount Sinai, New York City, NY, 10029, USA.

[§] **Corresponding author:** ST: windytian@hotmail.com; MSF: Mayte.SuarezFarinas@mssm.edu

# Abstract

**Background**
With the fast evolution of high-throughput technology, longitudinal gene expression experiments have become affordable and increasingly common in biomedical fields. Generalized estimating equation (GEE) approach is a widely used statistical method for the analysis of longitudinal data. Feature selection is imperative in longitudinal omics data analysis. Among a variety of existing feature selection methods, an embedded method — threshold gradient descent regularization (TGDR) stands out due to its excellent characteristics. An alignment of GEE with TGDR is a promising area for the purpose of identifying relevant markers that can explain the dynamic changes of outcomes across time.

**Methods**
We proposed a new novel feature selection algorithm for longitudinal outcomes— GEE-TGDR. In the GEE-TGDR method, the corresponding quasi-likelihood function of a GEE model is the objective function to be optimized and the optimization and feature selection are accomplished by the TGDR method.

**Results**
We applied the GEE-TGDR method a longitudinal lncRNA gene expression dataset that examined the treatment response of psoriasis patients to immune therapy. Under different working correlation structures, a list including 10 relevant lncRNAs were identified with a predictive accuracy of 80 % and meaningful biological interpretation.

**Conclusions**
A widespread application of the proposed GEE-TGDR method in omics data analysis is anticipated.

# Background

With fast evolution of high-throughput technology, longitudinal omics experiments have become affordable and increasingly common in many biomedical fields for exploring dynamically or temporally changed biological systems or processes. Usually, the analysis strategies focus on analyzing individual time points separately. As many investigators have pointed out [1–4], a failure to incorporate information contained in the dependent structure of time course data results in inefficient estimation of the standard errors, leading to an inadequate statistical power. Especially in big omics studies, this problem stands out since the sample size of such data is usually small. Furthermore, an over-simplified consideration by combining the results from marginal analysis at individual time points tends to fail to detect a meaningful pattern of changes over time.

The generalized estimating equation (GEE) approach [5] is a well-established and widely used statistical method to analyze longitudinal data. GEE considers the first two marginal moments (i.e., mean and variance) of data and a working correlation matrix to model correlated responses, artfully avoiding the specification of full joint likelihood function. The appeal of GEE lies in that it yields consistent estimators for the parameters of interest, even if the working correlation structure is incorrectly specified. Naturally, GEE has been modified or extended to identify differentially expressed genes over time for high-throughput data. Such modifications and/or extensions are not simple due to the high dimensionality of omics data, although some efforts have been made [1, 2, 6].

Like its cross-sectional counterpart, feature selection is imperative in the learning process for longitudinal omics data. Feature selection aims at eliminating irrelevant genes, avoiding over-fitting, speeding up the learning process and achieving a final model that is parsimonious (i.e., the number of selected genes is as least as possible). Consequently, a modification to GEE to analyze high-dimensional data necessitates the involvement of feature selection. In the literature, there are several such algorithms. For example, Wang et al [2] used an smoothly clipped absolute deviation (SCAD) penalty term [7] which is a novel extension to the $L_1$ penalty to equip the GEE models with feature selection capacity. The $L_1$ penalty, also known as LASSO [8], force genes with small estimated coefficients out of the final model, rendering a sparse solution by the means of which feature selection occurs. However, two subsequent works on this topic [1, 6] showed that this algorithm usually fails to converge when the number of covariates is much larger than the number of samples. This drawback is more apparent and fatal in longitudinal omics data, where the sample size is typically smaller than that of a cross-sectional study.

Among a variety of existent feature selection algorithms, we have devoted dramatic efforts on the threshold gradient descent regularization (TGDR) [9] method (see the Methods section for its description). Previously, we had extended TGDR for classification task of multiple groups (>2) and for identification of subgroup-specific prognostic genes with a survival outcome [10–14]. By applying these TGDR extensions to different types of omics data including microarray, RNA-Seq and mass spectrometry (MS) data, we have shown that TGDR and its respective extensions have many merits including easy-to-moderate programming intensity, good predictive performance and biologically meaningful implications of the resulting signatures. In a recent work [4], we show that the TGDR algorithm can be regarded as an optimization strategy, and that the final models given by TGDR have superior predictive performance and more meaningful biological interpretation than the LASSO models optimized by the coordinate descent method [15]. Therefore, we think an integration of GEE with TGDR may overcome the drawbacks of existing approaches for the purpose of longitudinal feature selection.

Long non-coding RNAs (lncRNAs) are post-transcriptional and epigenetic regulators, and have the characteristics of lower expression levels and more tissue-specific compared with protein-coding genes [16]. Once being regarded as evolutionary junks, lncRNAs have been demonstrated to play essential roles in many complex diseases, especially in cancers [16].

In this article, we proposed a new feature selection algorithm, referred to as GEE-TGDR, specifically for longitudinal data mining and feature selection. In the GEE-TGDR method, the corresponding quasi-likelihood function of a GEE model is the objective function to be optimized, while the optimization and feature selection are accomplished by the TGDR method. We applied this method to a longitudinal microarray gene expression data that aims at assessing the treatment efficacy of two immune therapies for psoriasis patients, and identified the relevant lncRNAs that can predict the temporal changes of psoriasis area and severity index (PASI) scores that is utilized to determine if a patient with psoriasis responds to the treatments, with the objectives of revealing the underlying mechanisms of these two treatments from the perspective of lncRNAs.

As pointed out by our previous study [17], psoriasis is an ideal model for examining the effects of targeted immune treatments given that it is well characterized by molecular profiles, displays low placebo effects and possesses easily accessible diseased tissues. So far, the implication of lncRNAs in psoriasis remains largely unexplored and poorly understood. Among the limited research carried out to explore the roles of lncRNAs play in psoriasis, however, some encouraging results have turned up. For example, a very recent study [18] has shown that antisense non coding RNA in the INK4 locus (ANRIL) can be regarded as a risk locus of psoriasis. Therefore, the roles of lncRNAs play in psoriasis deserve to be explore deeply and widely.

## Materials and Methods
*Experimental Data*
The microarray dataset [17] used in this study to validate the proposed GEE-TGDR algorithm was in the Gene Expression Omnibus (GEO) database (https://www.ncbi.nlm.nih.gov/geo/) under the accession number of GSE85034. There were 179 arrays in this experiment, including the gene expression profiles of 30 patients with moderate to severe psoriasis at the baseline non-lesion skins, baseline lesion skins, and weeks 1, 2, 4 and 16. Of the 30 patients, half were administrated with adalimumab (ADA) and the other half were treated with methotrexate (MTX). Of note, one patient on the ADA arm had no expression measurements of week 16 since his/her psoriasis area and severity index (PASI) score already had experienced a 75 % decrement at the week 4. In original paper, a treatment response was based on a reduction of 75 % in PASI score after week 12 or later. Longitudinal profiles of PASI scores (at baseline lesion skins, weeks 1, 2 and 4) were the outcomes of interest, and the lncRNA expression values of the baseline lesional skins serve as potential predictors to investigate if they are relevant to the PASI scores of psoriasis patients over time.

In this study, the pre-processed data were directly downloaded from the GEO database. No alternative pre-processing had been carried out. By matching the gene symbols of lncRNAs in the GENCODE (https://www.gencodegenes.org/) database (version 32) to those of genes annotated by the Illumina HumanHT-12 V 4.0 bead chips, 662 unique lncRNAs were identified and included in the downstream analysis.

*Statistical Methods*

In this paper, we conceive a new novel feature selection algorithm called GEE-TGDR specifically for selecting relevant features associated with the temporal changes of longitudinal outcomes, in which GEE is equipped with TGDR just as its name implies. We briefly described both GEE and TGDR methods before proceeding to the proposed integration. Here, to keep it the most relevant, we focused on the case of continuous outcomes.

*TGDR*

For continuous outcomes, the TGDR algorithm is based on a linear model, where a response variable $Y_i$ (i=1,…, n, n is the sample size) is modelled by a P-dimensional vector of observed covariates $X_{ip}$ (here, p =1, ..., P) as $E(Y)=X^T\beta$. Here, β's represents the coefficients of covariates for the magnitudes of association between covariates and the outcome. For continuous outcomes, a normal distribution is usually assumed and then the corresponding likelihood function is used as a response function/an objective function in the TGDR algorithm. With some algebraic simplification, the response function can be written as,

$$Res(\beta) = n^{-1} \sum_{i=1}^{n} (Y_i - X_i^T \beta)^2$$

The TDGR algorithm started from that the βs were initially set at zero's (corresponding to the null model). Using $\Delta v$ to denote a small positive increment (e.g., 0.01) in the gradient descent search and $v_k = k \times \Delta v$ to be the index for the point along parameter path for iteration k,

1. Upon current estimate $\beta(v_k)$, a negative gradient matrix g with its $p^{th}$ component as $g_p(v_k)$ are calculated as,

$$-\frac{\partial Res(\beta(v_k))}{\partial \beta_p} = g_p(v_k) = n^{-1} \sum_{i=1}^{n} x_{ip}(y_i - x_i^T \beta(v_k))$$

2. Let $f(v_k)$ represent the threshold vector of size p at iteration k, then its $p^{th}$ component (for the $p^{th}$ gene) is,

$$f_p(v_k) = I(|g_p(v_k)| \geq \tau \times \max (\underbrace{|g_l(v_k)|}_{l \in \beta}))$$

3. Update $\beta_p(v_{k+1}) = \beta_{jp}(v_k) - \Delta v \times g_p(v_k) \times f_p(v_k)$ and $v_{k+1} = v_k + \Delta v$.

4. Repeat steps 1-3 for K times. K can be regarded as a tuning parameter, with a large value corresponding to a dense model (more non-zero coefficients) and a small value to a sparse model (less non-zero coefficients). The optimal value of K is determined by cross-validations (CVs).

In the TGDR method, no explicit penalty term is added to the objective function (i.e., response function). The regularization on coefficients (thus the selection of features) is made possible by introducing the threshold function $f(v_k)$ in step 2, which determines if the gradient of a coefficient is large enough to descent or more precisely speaking to be updated. For more detailed description of the TGDR method, the works [9, 19] are referred.

## GEE

In the longitudinal notation, the $j^{th}$ time point/measurement of the $i^{th}$ subject, a t-dimensional vector of response variables $Y_{ij}$ (here, i =1, ..., n and j =1, ..., t) and a P-dimensional vector of covariates $X_{ip}$ (here, p =1, ..., P) are observed. Thus $Y_{i.} = (Y_{i1},...,Y_{it})^T$ denotes the vector of responses at t different time points for subject $i$ and $X_i = (X_{i1},...,X_{iP})^T$ is P covariates for subject $i$. For the purpose of a simplified notation, t is assumed to be the same for each individual.

In the GEE model, the first two marginal moments of $Y_{ij}$ are denoted by $\mu_{ij}(\beta)=E(Y_{ij})$ (the expectation of Y) and $\sigma^2(\beta) = V(Y_{ij})$ (the variance of Y). Here, β's are the coefficients representing the magnitude of association between covariates and outcomes, with $\beta_{jp}$ representing how attribute *p* is associated with the value of outcome $Y_{.j}$ (meaning the outcome at time point *j*). Those β's are parameters of interest. Furthermore, the distribution of $Y_{ij}$ is assumed to belong to an exponential family with a canonical link function. Let $\mu_i(\beta) = (\mu_{i1}(\beta_1),..., \mu_{it}(\beta_t))^T$ and $A_i(\beta) = \text{diag}(\sigma^2_{i1}(\beta_1),...,\sigma^2_{it}(\beta_t))$, then under a canonical link function $V_i(\beta) = A_i^{1/2}(\beta)R_i(\alpha)A_i^{1/2}(\beta)$. Here, $R_i(\alpha)$ is an t×t working correlation matrix with α as the finite dimensional parameter vector for correlations, which would be usually estimated by the residual-based moment method. In a GEE model, the quasi-likelihood function can be written as,

$$QL(\beta) = n^{-1} \sum_{i=1}^{n} (Y_{i.} - \mu_i(\beta))V_i^{-1}(\beta)(Y_{i.} - \mu_i(\beta))$$

Four structures are commonly used for the working correlation matrix $R_i(\alpha)$ — first-order autoregressive (AR1), exchangeable, unstructured and independent structure.

## GEE-TGDR

The conventional TGDR method only deals with univariate outcomes. As far as longitudinal outcomes that are multivariate are concerned, the method needs to be extended.

Here, we proposed to replace a likelihood function with a quasi-likelihood function and to extend TGDR as GEE-TGDR. With $\Delta v$ denoting a small positive increment (e.g., 0.01) in gradient descent search and $v_k = k \times \Delta v$ being the index for the point along parameter path after *k* steps, the GEE-TGDR algorithm is iterated in the following steps until the stopping criteria are met,

1. Upon current estimate $\beta(v_k)$, a negative gradient matrix g with its $(j,p)^{th}$ component as $g_{jp}(v_k)$ are calculated,

$$-\frac{\partial QL(\beta(v_k))}{\partial \beta_{jp}} = g_{ip}(v_k) = n^{-1} \sum_{i=1}^{n} X_{jp}^T A_i^{\frac{1}{2}}(\beta(v_k))R^{-1}(\alpha)A_i^{-\frac{1}{2}}(\beta(v_k))(Y_{i.} - \mu_i(\beta(v_k)))$$

2. Let $f_j(v_k)$ represent the threshold vector of size p for the $j^{th}$ time point (j=1,..,t) at step k, then its $p^{th}$ component (for the $p^{th}$ gene) is,

$$f_{jp}(v_k) = I(|g_{jp}(v_k)| \geq \tau \times \max_{l \in \beta_j}(|g_{jl}(v_k)|)) \; \forall j \in \beta_j$$

3. Update $\beta_{jp}(v_{k+1}) = \beta_{jp}(v_k) - \Delta v \times g_{jp}(v_k) \times f_{jp}(v_k)$ and $v_{k+1}$ to $v_k + \Delta v$.

4. Calculate the residuals $Y_i - \mu_i(\beta(v_k))$, and based on them, to estimate the nuisance parameters involved in $R(\alpha)$ (for different correlation structures, the parameters are different) and $A(\beta(v_k))$. Of note, since at different time points we have different threshold function, the selected genes at different time points are expected to differ. In this way, the selection of critical time points is possible.

5. Repeat steps 1-4 for K times. K is a tuning parameter, the same as in the conventional TGDR method. The optimal value of K is also determined by CVs.

In this study, we only developed the GEE-TGDR algorithm for continuous outcomes (given in the motivated database, PASI scores which are continuous were the outcomes of interest), thenthe corresponding expectations of Y's are,

$$E\begin{bmatrix} Y_{i1} \\ \vdots \\ Y_{it} \end{bmatrix} = \begin{bmatrix} \beta_{10} + \beta_{11}x_{i1} + \cdots + \beta_{1p}x_{ip} \\ \vdots \\ \beta_{t0} + \beta_{t1}x_{i1} + \cdots + \beta_{tp}x_{ip} \end{bmatrix}$$

Since the outcomes were continuous, the mean squared error (MSE) statistic was calculated to evaluate the performance of resulting gene signatures. It is worth pointing out that for the outcomes of other types, an extension suitable for the underlying data type of GEE-TGDR algorithm is straightforward, with the corresponding quasi-likelihood function serving as the objective function/response function.

*Statistical Language and Packages*
Statistical analysis was carried out in the R language version 3.6.1 (www.r-project.org).

## Results

In this study, we propose to extend the feature selection algorithm TDGR to account for correlation structure of longitudinal data. This is accomplished by defining the objective function of TDGR as the corresponding quasi-likelihood function, which as in GEE is specified based on the first two moments and a working correlation matrix. TDGR-GEE is described in the Materials and Methods section. In this section, we illustrate the application of the proposed method while looking for biomarkers that predict clinical resolution of psoriasis after being treated with two immune therapies.

Gene expression profiles of baseline lesional skin biopsies were obtained for 30 subjects followed up to 16 weeks after treatment with adalimumab and methotrexate. Clinical resolution at week 1, 2 and 4 was measured by PASI. In this example we would like to identify a signature of genes whose baseline expression valuess correlate with changes in PASI, our continuous longitudinal outcome. WE used 662 lnRNA as covariates in the proposed GEE-TGDR model, under 4 different working correlation structures. The performance statistics and identified lncRNA genes are presented in **Table 1**

In this example, the results obtained under working correlation structures exchangeable, unstructured and independent barely differ, with with similar set of biomarkers leading to similar performance. This reflects a well known robust characteristics of GEE, where when predictors are correctly given, the GEE estimates remain consistent even if the correlation structures is misspecified. Under the AR1 structure, GEE-TGDR identified only one lncRNA as being related to PASI scores, leading to an under-fitting and inferior to the performance when compared to the other three correlation structures.

Due to the patient burden and budgetary restrictions, longitudinal omics data are usually very short and unevenly spaced. In this case, AR1 is not well suited and the unstructured correlation may be the most suitable structure, even though that this structure corresponds to a model with more nuisance parameters involved in the corresponding working correlation structure.

Cross-validation results gave us an idea for the variability in the model performance in this regard, CV results indicated that all correlation structures but AR1 structure provided similar results, with both the exchangeable and independent structures having the least MSEs but a bigger variability and the unstructured structure having a larger MSE but the smaller variations.

Even though that at individual time points the identified features varied substantially for the unstructured, exchangeable and independent working correlation structures (**Figure1**), the unions of lncRNA lists across time are essentially the same, including 9 lncRNAs identified by all these three structures and one lncRNA selected by the independent structure alone (**Figure 2**).

In order to gain biological insight identified biomarkers, we evaluated the relevance to psoriasis of the 10 identified lncRNA using disease confidence scores, where a high score represents a solid support by the literature according to the GeneCards database. None of the 10 lncRNAs were directly related to psoriasis while 5 lncRNAs, listed in a descending order for the confidence scores, and thus descending support by the literature according to the GeneCards database, *MIR205*, *XIST, SNHG5, LINC01139* and *SDHAP2* were associated with immunity. Here, little meaningful information was extracted from currently annotated lncRNA databases, no surprisingly since that psoriasis remains largely unexplored from the perspective of lncRNAs. We thus focused on studying the mRNAs correlated or targeted by these lncRNAs. Specifically, we identified the genes whose baseline lesional expression was strongly correlated with at least one of the 10 lnRNA (|Spearman correlation coefficient|>0.6, FDR< 0.001) and identified 225 mRNAs genes. According to the GeneCards database[20], approximately 30 % of these mRNAs (64) were directly related to psoriasis, most notably *IL10, FABP5, KRT16, CCR6, IL18, STAT3, GATA3 and SERPINB3*, providing some validation of the lncRNA biomarkers identified by the GEE-TGDR method. In contrast, among the 29 target mRNAs identified by the lncRNA Disease 2.0 database [21] as targeted by the 10 lnRNA panel [all of which were identified by the correlation approach), GeneCards claimed that *CCR10, AOC3, UBB* and *WNK4* were directly related to psoriasis, but only *CCR10* had a large confidence score for its relevancy to psoriasis. Of note, among the 10 lncRNAs only *RAMP2-AS1, PAX1P1-AS1, TMEM99* and *LIN01018* have many correlated mRNAs, but the others five have few or no correlated mRNAs at all.

We conducted a gene-set overrepresentation analysis for the 225 mRNAs identified as targeted by the 10 lnRNA biomarker panel using the STRING software [22] on KEGG and GO collections. About 346 enriched biological process (BP) terms, 23 molecular function (MF) terms and 21 cellular component (CC) terms were identified in the GO collection reflecting the immune pathophysiology of the disease. The top 3 enriched KEGG pathways [23] reflected the inflammatory processes identifying *inflammatory bowel diseases* (FDR=0.0006) and *Cytokine-cytokine receptor interaction* (FDR=0.0051) but also zeroing on the hallmark pathway in psoriasis: *Th17 differentiation* (FDR=0.0313).

Lastly, among the 225 mRNA, we selected the top 10 in terms of psoriasis-relevance (confidence score for relevancy >15) and constructed a lncRNA-mRNA interaction network, visualized by Cytoscape software [24] (**Figure 4**). We observed that the target mRNAs are highly connected, with *IL10* serving as a hub gene. It is well-known that *IL10* is an immunosuppressive cytokine and enables to maintain immunological homeostasis [25]. Based on this, we anticipate that identified lncRNAs may regulate the expression of important cytokines such as *IL10* and warrant further investigation.

## Discussion

The GEE-TGDR method has several limitations. First, no grouping structure is taken into account and thus the GEE-TGDR method belongs to the conventional embedded feature selection category. So far, accumulated studies [26–29] have shown that a pathway-based method that considers grouping information is superior to its gene-based counterpart in which grouping information is ignored. Thus, how to extend the proposed GEE-TGDR method to account for correlations among genes is a research avenue we will pursue in the near future.

Second, the TGDR method is much slower than the coordinate descent (CD) [15] method as shown by our previous study [4] Given that the GEE-TGDR extension has the TGDR method as an optimization strategy, its speed of convergence is expected to be very slow. A method that combines the merits of these two algorithms together is definitely in demand.

Third, the GEE-TGDR method only takes time-invariant covariates in its current version. For longitudinal gene expression profiles, a summary score would be utilized to summarize each gene's gene expression values over time as one overall value. Consequently, covariates became time-constant again. For example, the mean values of lncRNA expression profiles at baseline and week 1 can be used to represent the corresponding lncRNAs and then as the covariates to investigate they are associated with PASI scores at week 1, week 2 and week 4 or the change of PASI scores at those time points from the baseline levels. On the other hand, the GEE-TGDR method can be certainly extended to handle time-varying covariates, which can examine the impact of dynamic changes in gene expression values on the outcomes of interest and thus facilitate a timely adjustment on treatment strategies accordingly.

Lastly, right now the only type of outcomes is continuous, yet certainly it can be extended to handle outcomes of other types, with the corresponding quasi-likelihood function acting as the objective function. Right now, over-fitting might be possible on the basis of the large discrepancy in MSE statistics between the whole training set and the cross-validations. Even worse but more realistic, over-fitting and under-fitting may accompany each other to exist in a feature selection process. Since for the real-world applications, the true relevant genes are unknown so the biological relevance is usually resorted to abstract some insight about the appropriation of identified gene lists. Nevertheless, for psoriasis and the underlying mechanism of immune treatments to combat this disease, little has been investigated from the perspective of lncRNAs to mine such relevant information. To the best of our knowledge, our work here is one of first efforts to unveil the mechanisms of psoriasis and its immune treatments using longitudinal lncRNA expression profiles and a feature selection method specific for such data.

With future work to eliminate these limitations, we believe that an lncRNA signature will be harvested to tell precisely which patients would respond to an immune treatment from those who would not, and thus facilitating personalized regimens or at least complementing other molecular markers for precise treatment strategies.

## Conclusions

Besides dealing with longitudinal clinical outcomes, the GEE-TGDR can be adopted to inference the associations between lncRNAs and mRNAs and thus construct lncRNA-mRNA interaction networks. For example, using well-known cancer-related mRNAs as outcomes, the lncRNAs that may potentially regulate/target those mRNAs could be found with the aid of the GEE-TGDR method. Therefore, we anticipate a widespread application of the proposed method in omics data analysis.

## Abbreviations
AR1: first-order autoregressive; GEE: generalized estimating equation; GEO: gene expression omnibus; MSE: mean squared error; CD: coordinate descent; PASI: psoriasis area and severity index; TGDR: threshold gradient descent regularization.

## Declaration

### Ethics statement and consent
Not applicable

### Consent to publish
Not applicable

### Availability of data and materials
Pre-processed gene expression data (Accession No: GSE85034) along with patient's clinical information were downloaded from the GEO database (https://www.ncbi.nlm.nih.gov/geo/ ).

### Competing interests
No competing interests have been declared.

### Funding
This study was supported by a fund (No. 31401123) from the National Natural Science Foundation of China and a fund (No. JJKH20190032KJ) from the Education Department of Jilin Province.

### Author's Contributions
Conceived and designed the study: ST. Analyzed the data: ST CW. Interpreted data analysis and results: CW ST. Wrote the paper: ST CW. All authors reviewed and approved the final manuscript.

### Acknowledgements
We thank Dr. Danna Gilbreath for the English editing.

**Table 1. Results of psoriasis lncRNA longitudinal data (using baseline expression values only as predictors)**

|  | Ave. of MSE (5-fold CVs) | SD of MSE (CVs) | MSE (all data) | Identified lncRNAs (using all data) | | | |
|---|---|---|---|---|---|---|---|
|  |  |  |  | Baseline | Week 1 | Week 2 | Week 4 |
| AR1 | 14.456 | 3.258 | 2.101 | RAMP2-AS1 | RAMP2-AS1 | RAMP2-AS1 | RAMP2-AS1 |
| Unstructured | 3.725 | 0.498 | 0.793 | XIST<br>RAMP2-AS1<br>MIR205 | LRRC75A-AS1<br>PAXIP1-AS1<br>LINC00667<br>RAMP2-AS1<br>MIR205 | LRRC75A-AS1<br>TMEM99<br>LINC01018<br>PAXIP1-AS1<br>LINC01139<br>RAMP2-AS1 | TMEM99<br>LINC01018<br>PAXIP1-AS1<br>LINC01139<br>RAMP2-AS1 |
| Exchangeable | 2.758 | 1.649 | 0.767 | XIST<br>RAMP2-AS1<br>MIR205 | LRRC75A-AS1<br>XIST<br>LINC01139<br>SDHAP2<br>RAMP2-AS1 | TMEM99<br>LINC01139<br>RAMP2-AS1 | TMEM99<br>XIST<br>LINC01018<br>PAXIP1-AS1<br>LINC01139<br>RAMP2-AS1 |
| Independent | 2.675 | 1.694 | 0.760 | SNHG5<br>LINC01139<br>RAMP2-AS1<br>MIR205 | SNHG5<br>RAMP2-AS1<br>MIR205 | SNHG5<br>TMEM99<br>RAMP2-AS1<br>MIR205 | SNHG5<br>XIST<br>LINC01018<br>LINC01139<br>RAMP2-AS1<br>MIR205 |

AR1: autoregressive order 1; MSE: mean squared error; SD: standard deviation; CV: cross-validation.

# Figures

Figure 1. Venn-diagram of identified lncRNAs for baseline, weeks 1, 2 and 4 respectively by different working correlation structures. **A)** Under the unstructured working correlation structure. **B)** The exchangeable working structure. **C)** The independent working structure.

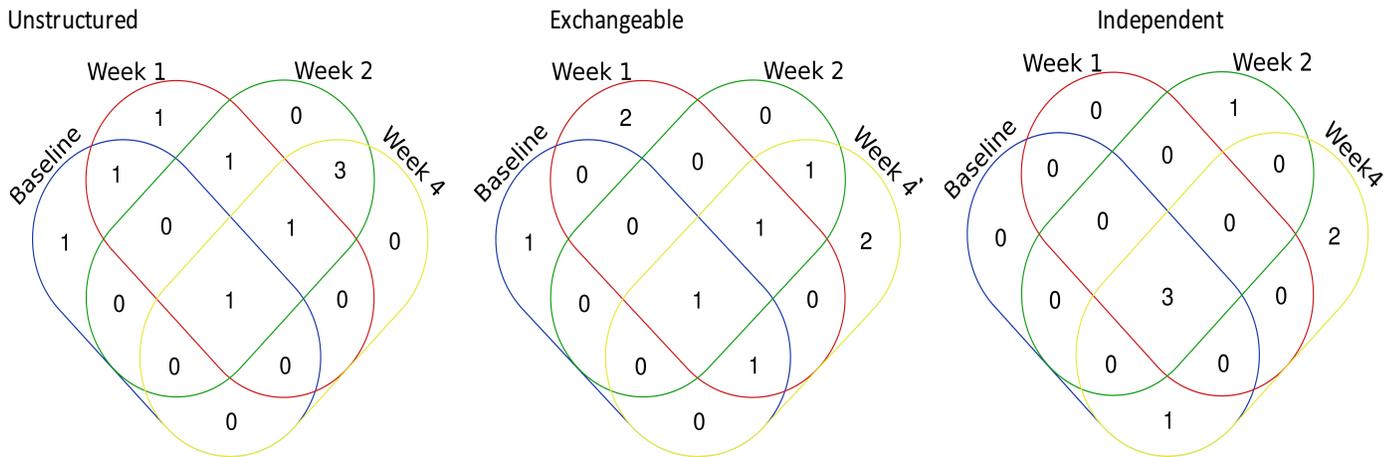

Figure 2. Venn-diagram of integrated lncRNAs by three working correlation structures.

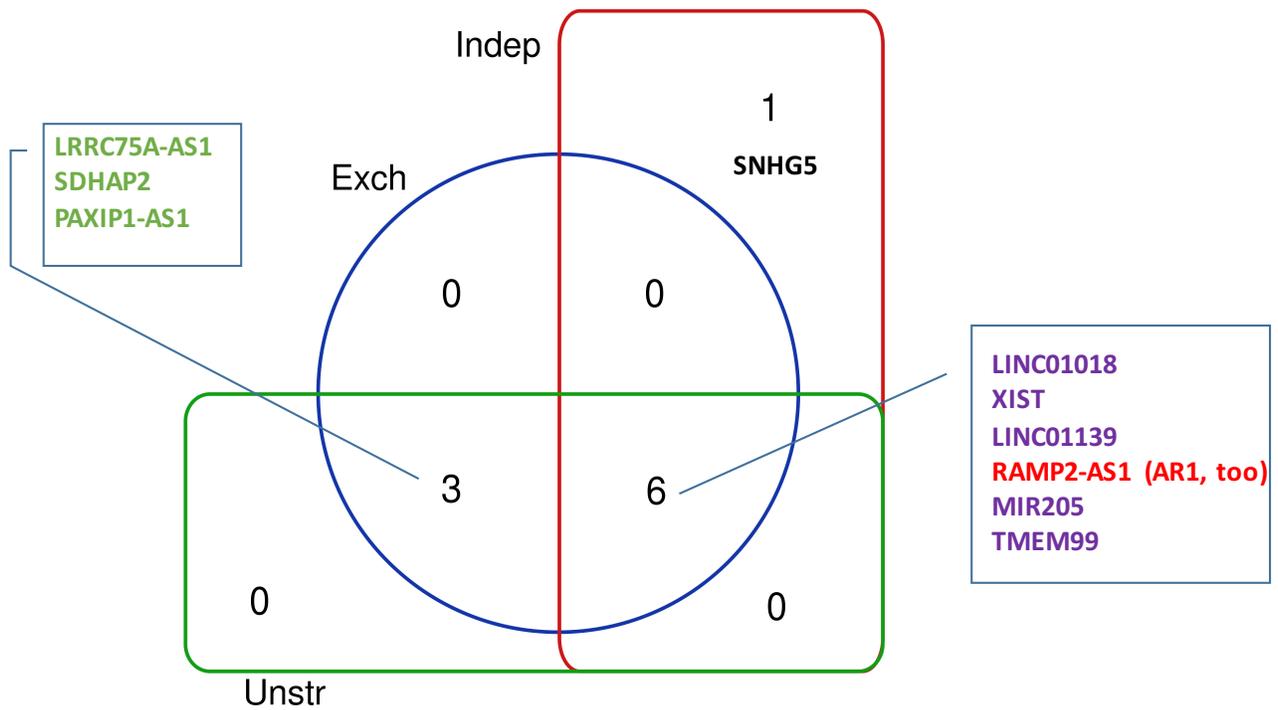

**Figure 3. Temporal trajectories of the identified 10 lncRNAs showing that all lncRNAs exhibited going back to normal level patterns.**

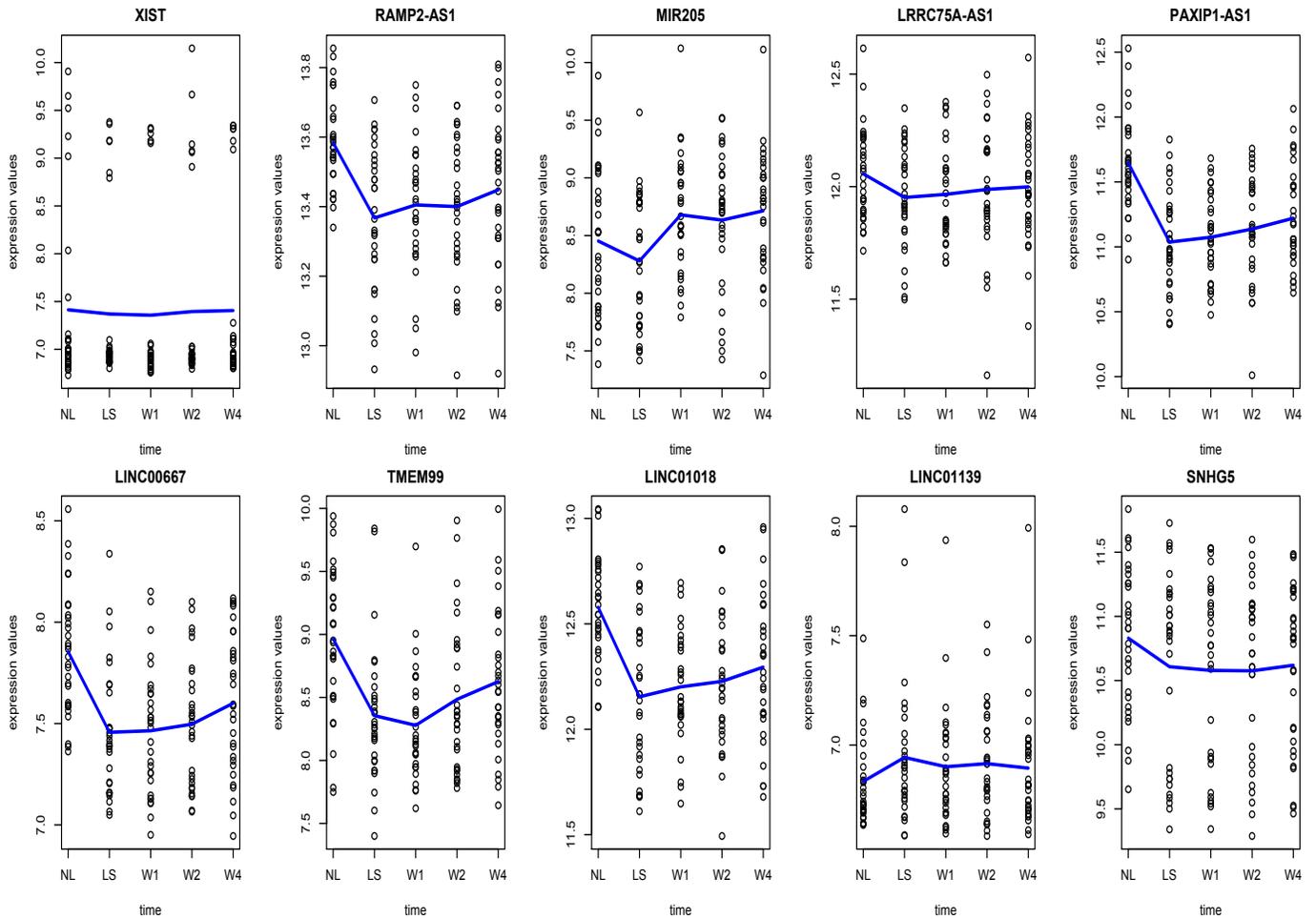

**Figure 4. Resulting interaction network of identified lncRNAs and their correlated mRNAs.** Here, only mRNAs with high enough confidence scores for the relevancy to psoriasis were considered. From the network, it is observed that IL10 is a hub gene directly connecting several other mRNAs and three identified lncRNAs. Four lncRNAs were highlighted in yellow, and the other six lncRNAs without correlated mRNAs were omitted from the graph.

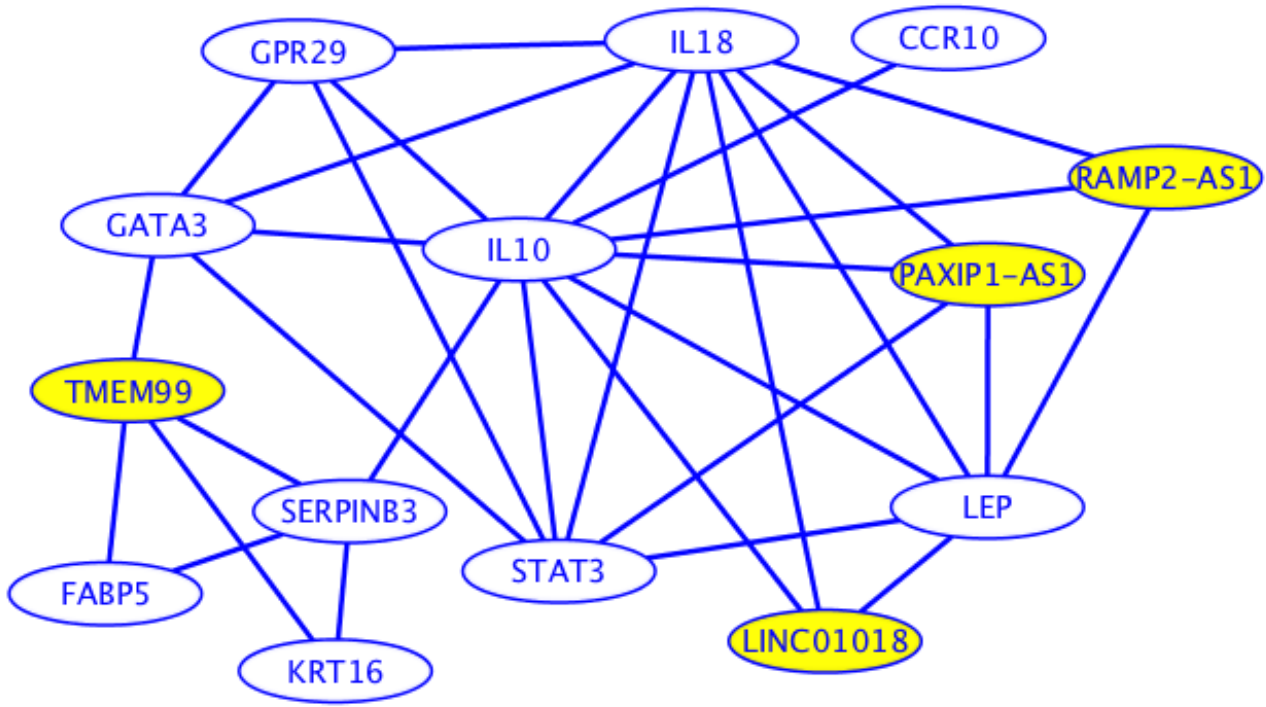